\documentclass{ws-ijgmmp}

\usepackage{mathtools}
\usepackage{tipa}
\usepackage{hyperref}

\newcommand{\be}{\begin{equation}}
\newcommand{\ee}{\end{equation}}
\newcommand{\ba}{\begin{eqnarray}}
\newcommand{\ea}{\end{eqnarray}}
\newcommand{\bs}{\begin{subequations}}
\newcommand{\es}{\end{subequations}}

\newcommand{\lp}{\left(}
\newcommand{\rp}{\right)}
\newcommand{\lb}{\left[}
\newcommand{\rb}{\right]}

\newcommand{\bbe}{\boldsymbol{\mathrm{e}}}

\newcommand{\bbie}{\boldsymbol{\textbf{\textschwa}}}

\newcommand{\+}{ \prescript{+}{}}
\newcommand{\m}{ \prescript{-}{}}
\newcommand{\volume}{\star 1}

\newcommand{\0}{0$^\text{th}$}
\newcommand{\1}{1$^\text{st}$}
\newcommand{\2}{2$^\text{nd}$}
\newcommand{\3}{3$^\text{rd}$}

\newcommand{\diff}{\text{d}}
\newcommand{\bdiff}{\textrm{\bf d}}
\newcommand{\bDiff}{\textrm{\bf D}}
\newcommand{\Diff}{\textrm{D}}

\newcommand{\bA}{\boldsymbol{A}}
\newcommand{\bB}{\boldsymbol{B}}
\newcommand{\bC}{\boldsymbol{C}}

\newcommand{\bF}{\boldsymbol{F}}

\newcommand{\bM}{\boldsymbol{M}}

\newcommand{\bK}{\boldsymbol{K}}
\newcommand{\bL}{\boldsymbol{L}}
\newcommand{\bT}{\boldsymbol{T}}

\newcommand{\bO}{\boldsymbol{O}}

\newcommand{\bR}{\boldsymbol{R}}
\newcommand{\bX}{\boldsymbol{X}}

\newcommand{\bp}{\boldsymbol{p}}
\newcommand{\bq}{\boldsymbol{q}}

\newcommand{\bt}{\boldsymbol{t}}

\newcommand{\bomega}{\boldsymbol{\omega}}

\newcommand{\bkappa}{\boldsymbol{\kappa}}

\begin{document}

\markboth{Authors' Names}
{Instructions for Typing Manuscripts (Paper's Title)}


%
\catchline{}{}{}{}{}
%

\title{Cosmology in the Lorentz gauge theory}

\author{Tomi Koivisto}

\address{Laboratory of Theoretical Physics, Institute of Physics, University of Tartu, W. Ostwaldi 1, 50411 Tartu, Estonia}
\address{National Institute of Chemical Physics and Biophysics, R\"avala pst. 10, 10143 Tallinn, Estonia}

\maketitle


\begin{abstract}

This proceeding is an introduction to cosmological applications of the Lorentz gauge theory. 
It provides the ingredients for a unique, though yet tentative $\Lambda$CDM theory of cosmology. 
Emergence of spacetime is described by the spontaneous symmetry breaking called 
here the khronogenesis. Space is then associated with the field strength of the antiself-dual gauge potential, and 
gravity is associated with the self-dual field strength. In the cosmological setting, khronogenesis
seems to predict inflation. It is shown that the Lorentz gauge theory allows the consistent description of
spin currents which could have important roles in cosmological phenomenology. 
 
\end{abstract}



\tableofcontents


\section{Introduction}	
 
In a discussion at the workshop Metric Affine Frameworks for Gravity at Tartu 2022 it was suggested that in a fundamental theory of spacetime and gravitation we should not presuppose a Metric but instead describe its emergence in terms of more elementary objects, whilst gauge theories based on symmetries under Affine transformations  
are known to describe the material dynamics on Hamiltonian lattice and continuum fields {\it in} spacetime yet may not furnish the most elementary Framework to describe the dynamics {\it of} spacetime, Gravity. 

This proceeding is an introduction to the Lorentz gauge theory, a new {\it pregeometric} framework for gravity and cosmology \cite{Zlosnik:2018qvg}. 
In pregeometric theories, the metric can arise as a composite object constructed from fundamental fermions \cite{Akama:1978pg,Amati:1981rf,Wetterich:2003wr,Wetterich:2021cyp}. Such a theory can be formulated without any reference to the metric of spacetime, and accommodate the ground state wherein the metric vanishes. Thus, there exists a ground state that describes the absence of spacetime rather than a given reference spacetime. In contrast to conventional gauge theories of gravity (see e.g. \cite{hehl} for some of the seminal papers), the Lorentz gauge theory is not formulated on an affine bundle, but is founded on a different approach akin to parameterised field theories \cite{Koivisto:2022uvd}.  

The theory introduced by Z\l{}o\'snik {\it et al} is based on the complexified Lorentz group. 
It is not necessary to dwell on the foundational importance of Lorentz symmetry. However, the complexification demands some justification from physics and not only from mathematical convenience. The justification is chirality. Matter, from macroscopic objects (like us) to its most elementary constituents (Weyl fermions) has chiral features, and we may ask should not the spacetime and the gravitational interaction reflect this property of matters. It can be incorporated by complexification, when working in tensor representations. Having the Lorentz-covariant derivative operator $\bDiff$ we can, in the complexified theory, introduce its self-dual and antiself-dual projections $\prescript{\pm}{}\bDiff$. In the spinor representation, these are just the projections that act on the left- and the right-handed Weyl spinors, respectively\footnote{This is despite the technicality that in (\ref{action}), following \cite{Ashtekar:1989ju}, it is assumed that the $\bL_M$ involves derivatives of only the conjugate of the right-handed spinor.}. (The definitions of the projections are recalled below in \ref{chiral}, and the exterior algebra notation below in \ref{exterior}.)
     
Now, we can deduce the action. A field needs to be introduced into the theory, since only topological invariants can be constructed from the operator $\bDiff$ alone. 
We consider a field $\phi^a$ in the fundamental representation. Again, the more fundamental formulation would take place in the spinor representation s.t. $\phi^a = \bar{\psi}\gamma^a\psi$, but working with tensors is convenient and suffices for our present purposes. Demanding that the shift symmetry $\phi^a \rightarrow \phi^a + \chi^a$ under global translations $\bDiff\chi^a=0$ is preserved, the action is determined 
\be \label{action}
I = \frac{i}{2}\int \phi_a\bDiff\+(\bDiff\bDiff)\bDiff\phi^a + \int \bL_{\text{M}}(\bDiff\phi^a,\psi,\+\bDiff\psi)\,,
\ee
up to a boundary term that can be fixed by matching the conserved charges with the observables \cite{BeltranJimenez:2021kpj} (and of course up to a specification of the matter source $\bL_M$ we have included for generality). It turns out that the $\phi^a$ plays the role of a clock field, for which reason it was dubbed the khronon. When the clock doesn't tick, $\bDiff\phi^a=0$, time doesn't flow. Due to the global symmetry we demanded, this is gauge-equivalent to $\phi=0$, and it is this trivial solution we identify as the pregeometric ground state. It has been often contemplated if a universe could appear {\it ex nihilo}, where the {\it nihil} might mean a quantum field theory vacuum, a spaceless or a boundary-free geometry, or something else, e.g. \cite{tryon,Vilenkin:1982de,Hartle:1983ai,Gott:1997pm}. Since the pregeometric ground state in our theory offers a candidate for the ``nothing'', a question we shall begin to explore is whether the action principle (\ref{action}) alone might determine (at least some of) the boundary conditions for the universe, and in particular, explain its (hypothetically inflationary) beginnings. 

Thus, the focus of this proceeding is on cosmology. The main results of the following sections, from the perspective of cosmology, are listed below.
\begin{itemize}
\item In section 2 we introduce the khronogenesis, emergence of space and time via a spontaneous Lorentz symmetry breaking. 
\item Section 3 shows that theory (\ref{action}) is cosmologically viable without any dark matters in $\bL_M$, and points out a duality relating the $\Lambda$ and the CDM.
\item In section 4 we discover that the ground state can spontaneously yet continuously begin to inflate into our hot big bang universe. 
\item Section 5 introduces spin currents. They could be significant in early universe phenomenology, and potentially resolve the $H_0$ tension. 
\end{itemize} 
The only new (very simple) solutions are in sections 4 and 5. Many of the derivations and clarifications in the preceding sections haven't been published elsewhere either. In the final section 6 we point out some of the calculations that should be tackled next, and discuss some of the new possibilities for cosmological model-building. 
 

\subsection{Lorentz algebra}
\label{chiral}

The algebra $\mathfrak{so}(4,\mathbb{C})$ has 2 invariants: $\eta_{ab}$ and $\epsilon_{abcd}$. Our convention is $\eta_{ab}=(-1,1,1,1)$, and $\epsilon_{0123}=1$.
Consider a bivector $X^{ab}$ in the algebra. Its $\star$-dual is defined as
\be
\star X^{ab} = \frac{1}{2}\epsilon^{ab}{}_{cd}X^{cd} \quad \Rightarrow \quad \star\star X^{ab} = -X^{ab}\,. 
\ee
We can also define the (anti)self-dual projections,
\be \label{projectors}
\prescript{\pm}{}X^{ab} = \frac{1}{2}\lp 1 \mp i\star\rp X^{ab}  = \frac{1}{2}\lp \delta^a_c\delta^d_b \mp \frac{i}{2}\epsilon^{ab}{}_{cd}\rp X^{cd} \quad \Rightarrow \quad \star \prescript{\pm}{}X^{ab} = \pm i  \prescript{\pm}{}X^{ab}\,. 
\ee 
These are indeed projections since
\bs
\ba
X^{ab} & = & \+ X^{ab}  + \prescript{-}{}X^{ab} \,, \\
\prescript{\pm}{} (\prescript{\pm}{}X^{ab} ) & = & \prescript{\pm}{}X^{ab}\,, \\
\prescript{\mp}{} (\prescript{\pm}{}X^{ab} ) & = & 0\,.
\ea
\es
It is useful to note properties of the products
\bs
\ba
\prescript{\pm}{}X^{ab}  Y_{ab}  & = & X^{ab}  \prescript{\pm}{}Y_{ab}  =  \prescript{\pm}{}X^{ab}  \prescript{\pm}{}Y_{ab} \,,  \label{product1}  \\
\prescript{\pm}{}X^{ab}   \prescript{\mp}{}Y_{ab}  & = & 0\,,  \label{product2}  \\
\epsilon_{abcd}\prescript{\pm}{}X^{ab}Y^{cd} & = & \epsilon_{abcd}X^{ab}\prescript{\pm}{}Y^{cd} =  \epsilon_{abcd}\prescript{\pm}{}X^{ab}\prescript{\pm}{}Y^{cd}\,, \label{product3} \\
 \epsilon_{abcd}\prescript{\pm}{}X^{ab}\prescript{\mp}{}Y^{cd} & = &  0\,,  \label{product4} 
\ea
\es
which follow immediately from the definition of the projection. 
The $\pm$ split basically realises the isomorphism $\mathfrak{so}(4,\mathbb{C}) = \mathfrak{su}(2,\mathbb{C})\times\mathfrak{su}(2,\mathbb{C})$. In the latter form, we have 2 decoupled algebras with the 2 invariants: $\delta_{IJ}$ and $\epsilon_{IJK}$, where the indices $I$, $J$, $K$ take the values 1,2,3. We will be also using this form of the theory.

\subsection{Exterior algebra}
\label{exterior}

We denote $p$-forms with bold symbols if $p>0$. The antisymmetric wedge product is used explicitly. For a $p$-form $\bp$ and a $q$-form $\bq$, we have $\bp\wedge\bq = (-1)^{qp}\bq\wedge\bp$. To set up a Lorentz gauge theory, we introduce the Lorentz gauge potential 1-form $\bomega^{ab}$, implicit in the $\mathfrak{so}(4,\mathbb{C})$-covariant exterior derivative $\bDiff$. We note that $\bDiff(\bp\wedge\bq) = \bDiff\bp\wedge\bq + (-1)^p\bp\wedge\bDiff\bq$.  
We can perform the $\pm$ decomposition of the gauge potential $\bomega^{ab}=\+\bomega^{ab} + \m\bomega^{ab}$. The exterior derivative $\+\bDiff$ is then covariant only wrt self-dual Lorentz transformations, and the $\prescript{-}{}\bDiff$ is covariant only wrt to antiself-dual transformations. However, the projection of bivectors is Lorentz-invariant, since the symbols $\eta_{ab}$ and $\epsilon_{abcd}$ are. In particular, 
the field strength of the gauge potential,
\be \label{curvature}
\bR^{ab} = \bdiff\bomega^{ab} + \bomega^a{}_b\wedge\bomega^{bc} = \+\bR^{ab} + \m\bR^{ab}\,,
\ee
can be split into the 2 projections which are the field strengths of the respective 2 projections of the gauge potentials (only after we have moved towards spacetime geometry in section \ref{kinematics} we begin refer to the Lorentz gauge potentials also as connections). As one quickly checks using the Poincar{\'e} lemma $\bdiff^2=0$, we have $\bDiff^2 X^a=\bR^a{}_b X^b$ for a Lorentz vector $X^a$. Similarly, for a bivector $X^{ab}$ we obtain $\bDiff^2 X^{ab}=-2\bR^{[a}{}_c X^{b]c}$,
and if $\bX^{ab}$ is a bivector $p$-form, we write $\bDiff^2 \bX^{ab}=-2\bR^{[a}{}_c\wedge\bX^{b]c}$, etc. A conventional tool in spacetime geometry is the coframe $\bbe^a$. We define the 4-volume element as
\bs
\be
\star 1 = \frac{1}{4!}\epsilon_{abcd}\bbe^a\wedge\bbe^b\wedge\bbe^c\wedge\bbe^d\,,
\ee
and also the 3-form basis 
\be
\star\bbe^a =  \frac{1}{3!}\epsilon^a{}_{bcd}\bbe^b\wedge\bbe^c\wedge\bbe^d\,,
\ee
\es
will be useful. The 2-form basis $\bbe^{a}\wedge\bbe^b$ is a bivector (since it is antisymmetric) and therefore the rules in \ref{chiral} apply to it as well. It is now straightforward to show that e.g. $\bbe^a\wedge\star\bbe^b=-\eta^{ab}\star 1$, and $\bbe^a\wedge\bbe^b\wedge\star(\bbe_c\wedge\bbe_d) = - 2\delta^{[a}_c\delta^{b]}_d\star 1$. 
Given a spacetime geometry we could write the gravitational part of the action (\ref{action}) as an integral 
$I_G = \int \mathcal{L}_G\diff^4 x$ over a Lagrangian density $\mathcal{L}_G = (i\det{\bbe}/2)\epsilon^{\alpha\beta\gamma\delta}\phi_a\Diff_\alpha\+\Diff_\beta\+\Diff_\gamma\Diff_\delta\phi^a$.

\section{Lorentz gauge theory}

We begin the study of the Lorentz gauge theory (\ref{action}) by stating its EoM's (equations of motion) in \ref{fieldeqs}.
The spontaneous symmetry breaking giving rise to space and time is demonstrated straightaway in \ref{genesis}. 
The simple example of flat Minkowski spacetime is already a non-trivial Lorentz gauge field configuration with dynamical 
field strength.  
In \ref{kinematics} the structure of the theory is clarified by establishing the relations of the Lorentz gauge field strengths and 
more conventional geometrical objects, such as the curvatures of the metrical Levi-Civita connection or the self-dual Ashtekar connection. 
In \ref{dynamics} the dynamical equations are put into a convenient 1+3 form.

\subsection{Field equations}
\label{fieldeqs}

For generality, we have included matter sources for some fields $\psi$ in the action (\ref{action}). The variation of lagrangian 4-form $\bL_{\text{M}}$ wrt the field $\psi$ produces their Euler-Lagrange EoMs. 
The variations of $\bL_{\text{M}}$ wrt the fundamental fields can be parameterised in terms of the two 3-forms $\bt_a$ and $\bO^{ab}=\bO^{[ab]}$ as
\begin{subequations}
\label{mattervariations}
\ba
\frac{\delta \bL_{\text{M}}}{\delta \phi^a} & = & -\bDiff\bt_a\,, \\
\frac{\delta \bL_{\text{M}}}{\delta \bomega_{ab}} & = & - \phi^{[a}\bt^{b]} + \bO^{ab}\,.
\ea
\end{subequations}
The $\bt_a$ is the material energy-momentum 3-form and $\bO^{ab}$ is the material angular-momentum 3-form, which we may call more briefly the energy current and the spin current, respectively. They are the sources in the gravitational field equations we obtain from (\ref{action}),
\begin{subequations}
\label{efe1}
\ba
\bDiff(i\prescript{+}{}\bR^a{}_b\wedge \bDiff \phi^b - \bt^a) & = & 0\,, \label{efe1a} \\
\frac{i}{2}\bDiff \prescript{+}{}(\bDiff\phi^{[a}\wedge\bDiff\phi^{b]}) & = & i\phi^{[a}\+\bR^{b]}{}_c\wedge\bDiff\phi^c - \phi^{[a}\bt^{b]} + \bO^{ab}\,. \label{efe1b}
\ea
\end{subequations}
The \1 equation shows a covariantly closed 3-form. Let us just call this 3-form $\bM^a$,
\be \label{solution}
i\prescript{+}{}\bR^a{}_b\wedge \bDiff \phi^b - \bt^a = \bM^a \quad \text{where} \quad \bDiff \bM^a = 0\,. 
\ee
Using this in the \2 field equation (\ref{efe1b}) the system is 
\begin{subequations}
\label{efe2}
\ba
i\prescript{+}{}\bR^a{}_b\wedge \bDiff \phi^b & = & \bt^a + \bM^a\,,  \label{efe2a} \\
\frac{i}{2}\bDiff \prescript{+}{}(\bDiff\phi^{[a}\wedge\bDiff\phi^{b]}) & = & \phi^{[a}\bM^{b]} + \bO^{ab}\,. \label{efe2b}
\ea
\end{subequations}
The \1 of these equations may look familiar. The LHS would become the Einstein 3-form if we could identify (up to $i$) the self-dual field strength with a metric curvature 2-form, and identify the 1-form $\bDiff\phi^a$ with the coframe of a metric tensor $g$ s.t. $g = \eta_{ab}\bDiff\phi^a\otimes\bDiff\phi^b$. 

\subsection{Khronogenesis}
\label{genesis}

The simplest solution to the theory (\ref{action}) without matter sources is $\phi^a=0$. The gauge potential $\bomega^{ab}$ is then completely arbitrary. Due to the existence of this totally symmetric solution of the theory, it can be regarded as a pre-geometric theory of gravity and spacetime. Neither of these is postulated a priori, but they can emerge in a symmetry-broken phase of the theory.

Let us assume that $\phi^a\phi_a <0$. A time-like expectation value of the field breaks the Lorentz symmetry down to the rotational symmetry. For convenience, we may then adopt the gauge 
$\phi^a = \phi(x)\delta^a_0$ wherein the \0 axis is aligned with the field. It will turn out that the \0 component $\phi$ can then interpreted as a clock function, for which reason the field $\phi^a$ is called the khronon scalar. 

To see how time and space are constructed in such a symmetry-broken phase, let us first consider the most basic case, the flat-metric spacetime which is the background usually postulated in standard quantum field theory. For simplicity, we pick coordinates s.t. the clock function $\phi=t$ is the time coordinate. Then $\bDiff\phi^0 = \bdiff\phi = \bdiff t$, so the time component of the Minkowski coframe is reproduced correctly. The spatial components of the coframe, $\bDiff\phi^I = \bdiff \phi^I + \bomega^I{}_0\phi^0 = \bomega^I{}_0 t$ are now proportional to the electric components of the gauge potential. The spatial coframe of the Minkowski space would be $\bdiff x^I$. Thus we require the non-vanishing gauge potential $\bomega^I{}_0 = t^{-1}\bdiff x^I$. Is this a pure gauge potential? 

To compute the field strength of the potential, we need also its magnetic components which are not determined by the background geometry alone. The magnetic components $\bomega^I{}_J$ we can solve from the field equations. We choose the integration form $\bM^a=0$ to vanish at (\ref{solution}) and assume no sources $\bt^a=0$. Then (\ref{efe2a}) reduces to $\prescript{+}{}\bR^a{}_b\wedge \bDiff \phi^b=0$, which is satisfied if the self-dual field strength vanishes. Thus, we can set the self-dual gauge potential to vanish. Then
$\star\bomega^{ab} = -i\bomega^{ab}$, implying that the magnetic components of the potential are given by the electric components as $\bomega^{IJ} = i\epsilon^{IJ}{}_K\bomega^K{}_0$. To summarise, the Minkowski vacuum is supported by the field configuration
\be \label{Mconfiguration}
\phi = t\,, \quad \bomega^I{}_0  =  t^{-1}\bdiff x^I\,, \quad \bomega^I{}_J = it^{-1}\epsilon^I{}_{JK}\bdiff x^K\,.    
\ee
This simple but non-trivial configuration is described in gauge-invariant terms by the field strength
\bs
\label{Mcurvature}
\ba
\bR^{I0} & = & t^{-2}\lp \bdiff t\wedge\bdiff x^I + i\epsilon^I{}_{JK}\bdiff x^J\wedge\bdiff x^K\rp\,, \\
\bR^{IJ} & = & t^{-2}\lp -i\epsilon^{IJ}{}_{K}\bdiff t\wedge\bdiff x^K + 2\bdiff x^I\wedge\bdiff x^J \rp\,.   
\ea
\es   
Space and time, even the case of Minkowski, requires the anti-selfdual gauge field strength. 
In this sense, spacetime in our theory not only has but is geometry. 

Reference frames in gravitational theories can be described in terms of tetrad components. Now
we may compose a coframe 1-form $\bbe^a$ from the derivative of the symmetry-breaking field 
$\bbe^a=\bDiff\phi^a$. If this represents an invertible matrix, there exist the quartet of vectors $\bbie_a$ s.t. 
$\bbie_a\lrcorner\bbe^b=\delta^b_a$, and these four vectors $\bbie_a$ can then play the role of tetrads.
They do this for example in the Minkowski configuration (\ref{Mconfiguration}), but the theory (\ref{action}) 
remains well defined in generic configurations wherein the coframe composed as $\bDiff\phi^a$ 
can be degenerate. A criterion for a configuration to describe a spacetime is that there exists a non-degenerate tetrad. 
Only then can we define conventional spacetime tensors such as e.g. the 4-index Riemann curvature tensor 
$\bbie_b\lrcorner(\bbie_a\lrcorner\bR^{ab})$.  

Another invariant characterisation of geometry is the torsion 2-form $\bT^a$ defined as the derivative of the 
coframe, $\bT^a = \bDiff\bbe^a$. In conventional models of gravity, torsion is independent of the curvature. However, now
we find the relation
\be \label{torsioncurvature}
\bT^a = \bDiff\bbe^a =  \bDiff\bDiff\phi^a = \bR^a{}_b\phi^b\,. 
\ee    
By taking further the derivative of this relation, we obtain $\bDiff\bT^a = \bR^a{}_b\wedge\bbe^b$, which
is a geometric Bianchi identity satisfied in any gravity theory, but the relation (\ref{torsioncurvature}) is peculiar to
the Lorentz gauge theory. For the Minkowski solution (\ref{Mconfiguration}) the relation implies that
\bs
\label{Mtorsion}
\ba
\bT^0 & = & 0\,, \\
\bT^I & = & t^{-1}\lp \bdiff t\wedge\bdiff x^I + i\epsilon^I{}_{JK}\bdiff x^J\wedge\bdiff x^K\rp\,.
\ea
\es
Both the curvature and the torsion of the Minkowski solution exist in the anti-self dual sector of the theory.
We have $\bR^{ab}=\m\bR^{ab}$, and also $\bT^a=\m\bT^{a}$. In the asymptotic future, the khronon grows
without a bound and the geometry (\ref{Mcurvature},\ref{Mtorsion}) fades away. There is no asymptotic past,
but a singularity at $\phi=t \rightarrow 0$. Thus, in fact this geometry could not have been created in 
khronogenesis. The lesson is that a transition ``from no space to flat space'' is impossible.

We have to study the theory in more depth and consider a bit more elaborated model.  A viable class of khronogenetic spacetimes 
will then be arrived at in \ref{isodyn}.

\subsection{Kinematics}
\label{kinematics}

It can be useful to look in more detail at the geometric structure of the theory and see how it gives rise to the standard description of general relativity in terms
of the $\bbe^I$-compatible torsion-free connection $\mathring{\bomega}^I{}_J=\epsilon^I{}_{JK}\mathring{\bomega}^K$,
and the extrinsic curvature 1-form ${\bK}_I= \bbie_I\lrcorner(\mathcal{L}_{\bbie_a\phi^a}h)$ computed from the canonical spatial metric $h=\delta_{IJ}\bbe^I\otimes\bbe^J{}$. 

It can be deduced that the connection in the time gauge assumes the form \cite{Zlosnik:2018qvg}
\be \label{omega}
{\bomega}^{ab} = 
\lp
\begin{matrix}
0 & \bbe^I/\phi  \\
-\bbe^I/\phi  & \epsilon^{IJ}{}_{K}\lp 2\bA^K + i\bbe^K/\phi\rp 
\end{matrix}
\rp\,, 
\ee
where we have denoted $\bA^I = \frac{1}{2}(\mathring{\bomega}^I - i\bK^I)$. The self-dual and the anti-self-dual parts of this connection are 
\begin{subequations}
\label{omega+-}
\ba
\+\bomega^{ab} & = & 
\lp
\begin{matrix}
0 & i \bA^I  \\
-i \bA^I  & \epsilon^{IJ}{}_{K}\bA^K
\end{matrix} \label{omega+}
\rp\,, \\ 
\prescript{-}{}\bomega^{ab} & = & 
\lp
\begin{matrix}
0 &  -i \bA^I  + \bbe^I/\phi \\
i \bA^I -  \bbe^I/\phi  & \epsilon^{IJ}{}_{K}\lp \bA^K + i\bbe^K/\phi\rp \label{omega-}
\end{matrix}
\rp\,. 
\ea
\end{subequations}
Then the curvature can be written as 
\be \label{Rab}
\bR^{ab} = 
\lp
\begin{matrix}
0 & \bT^I/\phi \\
-\bT^I/\phi  & \epsilon^{IJ}{}_{K}\lp 2\bF^K + i \bT^K/\phi \rp
\end{matrix}
\rp\,. 
\ee
We have denoted 
\begin{subequations}
\label{ashtekarcurvature}
\be \label{ashtekar}
\bF^I = \bdiff \bA^I - \epsilon^{I}{}_{JK}\bA^J\wedge\bA^K\,. 
\ee
The connection $\bA^I$ corresponds to the Ashtekar connection and $\bF^I$ is its curvature \cite{Ashtekar:1986yd,Peldan:1993hi}.  
The latter can also be expressed in terms of the metric curvature $\boldsymbol{\mathring{R}}{}^{IJ}$ and the metric-covariant derivative ${\mathring{\bDiff}}$ as
\be \label{levi-civita}
\bF^I = \frac{1}{4}\epsilon^{I}{}_{JK}\lp {\mathring{\bR}}{}^{JK} +\bK^J\wedge\bK^K\rp -\frac{i}{2}{\mathring{\bDiff}}\bK^I\,. 
\ee
\end{subequations}
In the decomposition of (\ref{Rab}), the self-dual part involves only these standard ingredients,   
\begin{subequations}
\label{Rab2}
\ba \label{dualcurv}
\+\bR^{ab} & = & 
\lp
\begin{matrix}
0 & i\bF^I  \\
-i\bF^I  & \epsilon^{IJ}{}_{K}\bF^K
\end{matrix}
\rp\,, \\
\prescript{-}{}\bR^{ab} & = & 
\lp
\begin{matrix}
0 &  - i\bF^I  + \bT^I/\phi \\
 i\bF^I - \bT^I/\phi  & \epsilon^{IJ}{}_{K}\lp \bF^K + i \bT^I/\phi \rp
\end{matrix}
\rp\,.
\ea
\end{subequations}
The expression for torsion, generalising (\ref{Mtorsion}) is,
\begin{subequations}
\label{torsionexp}
\ba
\bT^0 & = & 0\,, \\
\bT^I & = & \bdiff \bbe^I + \bbe^I \wedge \bdiff\log{\phi} - \epsilon^I{}_{JK}\lp 2\bA^J + i\bbe^J/\phi\rp\wedge\bbe^K\,. \label{torsionexpb} 
\ea
\end{subequations}
The presence of the torsion 2-form distinguishes the $\+\bR^a{}_b$ and $\prescript{-}{}\bR^a{}_b$ as independent of each other. 
 As we found in the previous subsection, the $\prescript{-}{}\bR^a{}_b$ sector is excited even in the flat Minkowski background spacetime. 
In this subsection, we have not used the EoM's but only looked at the kinematic structure.

\subsection{Dynamics}
\label{dynamics}

We have now considered two decompositions of the Lorentz gauge field strength. It can be split into the self-dual and antiself-dual field strengths $\prescript{\pm}{}\bR^{ab}$, but an alternative split in the time gauge was made in terms of the two Lorentz 3-vector 2-forms, the curvature $\bF^I$ and the torsion $\bT^I$, given by equations (\ref{ashtekarcurvature}) and (\ref{torsionexpb}), respectively.  Exploiting the latter decomposition, we find that the gauge-fixed action reduces to
\begin{subequations}
\be
I = 
\int \lb i\bbe^0\wedge\bbe^I + 2\star\lp\bbe^0\wedge\bbe^I\rp\rb\wedge\bF_I + \int \bL_M\,,
\ee
and can be further massaged into the remarkably simple form
\be
I = \frac{1}{2}\int \phi \bF_I\wedge\bT^I + \int\bL_M\,.
\ee
\end{subequations}
This looks like a $\mathfrak{su}(2,\mathbb{C})\times\mathfrak{su}(2,\mathbb{C})$ version of the theory. 
We can also rewrite the field equations (\ref{efe2}), as
\begin{subequations}
\label{summaryeqs}
\ba
\bF_I\wedge \bbe^I & = & \bM^0 + \bt^0\,, \label{curvatureeq1} \\
\bF^I\wedge\bbe^0 + i\epsilon^I{}_{JK}\bF^J\wedge\bbe^K & = & -\bM^I - \bt^I\,, \label{curvatureeq2} \\
\bT^I\wedge\bbe^0 - i\epsilon^I{}_{JK}\bT^J\wedge\bbe^K & = & i\phi\bM^I + 2i\bO^{0I}\,. \label{torsioneq}
\ea
The coframe field can be considered as the short-hand notation for $\bbe^0 = \bdiff\phi$, $\bbe^I = \phi\bomega^I{}_0$. 
Assuming that $\prescript{-}{}(\bO^{ab} + \phi^{[a}\bM^{b]})=0$, this is the full set of field equations. Otherwise, the \2 torsion equation,
\be
\bT^{[I}\wedge\bbe^{J]}+\frac{i}{2}\epsilon^{IJ}{}_K\bT^K\wedge\bbe^0 =  -2i\bO^{IJ}\,,
\ee
\end{subequations}
is not redundant with (\ref{torsioneq}), but their combination will impose the latter constraint.  

\section{The $\Lambda$-space $\bkappa$ and the CDM-time $\phi$}

In general, we have deduced that the 3-form $\bM_a$ obeys 
\bs
\label{summarycdm}
\ba
\bDiff \bM_a & = & 0\,, \label{cdma} \\
\bM_K & = & -\phi^{-1}\lp 2\bO^0{}_K - i\epsilon_{IJK}\bO^{IJ}\rp\,. \label{cdmb}
\ea
\es 
In the next subsection \ref{thecdm} we will solve these equations and show that cosmology in the Lorentz gauge theory can
be viable without dark matter. Then, in \ref{lambda} we introduce another 3-form $\bkappa$ conjugated to a cosmological
constant, and discover a cosmic duality between the ``local time''  measured by $\phi^a$ and the  
``global time''  measured by $\bkappa$ \cite{Gallagher:2021tgx}.

\subsection{The CDM}
\label{thecdm}

The properties of the 3-form $\bM_a$ can be deduced from its EoM (\ref{summarycdm}) in 3 steps. 

\subsubsection{The $\bM^a$ is aligned with the khronon $\phi^a$.}
\label{aligned}

To begin with, in (\ref{action}) it is assumed that material sources do not have non-minimal couplings to the antiself-dual connection. 
Then equation (\ref{cdmb}) dictates that $\bM_I=0$, so that $\bM^a$ as a Lorentz vector has only the possible non-zero 
time-like component. We could deduce this without gauge-fixing by starting from (\ref{efe2b}), and by setting the non-minimal
matter coupling to vanish, $\m\bO^{ab}=0$. Then it follows that also $\m\phi^{[a}\bM^{b]}=0$, from which it follows that
$\phi^{[a}\bM^{b]}=0$, so that $\bM^a \sim \phi^a$, which in the time gauge is just the statement $\bM_I=0$. 

The point is that we could relax the assumptions $\m\bO^{ab}=0$ and $\phi^{[a}\bM^{b]}=0$ only together but not separately.
The antiself-dual matter hypermomentum and the geometrical spin current must cancel each other, if they are non-zero.
This might be an interesting possibility to explore, but in these proceedings we stick to the assumption $\m\bO^{ab}=0$.

\subsubsection{The $\bM^a$ is a 3-space volume form.}

We have now reduced the four 3-form components of $\bM^a$ to just one 3-form $\bM^0$. As it arises from the integration of the gravitational
field equations, a priori we would have to specify its four independent components in order to evolve the dynamical system. However,
upon closer look at the structure of the theory, the number of the required initial conditions can be further reduced. Now, the spatial components of the
conservation equation (\ref{cdma}) give
\be
\bDiff\bM_I = \bdiff\bM_I - \bomega^0{}_I\wedge\bM_0 - \bomega^J{}_I\wedge\bM_J = -\phi^{-1}\bbe_I\wedge\bM_0 = 0\,, 
\ee  
where we used the previous step $\bM_I=0$ and the definition of the composite coframe $\bbe^a = \bDiff\phi^a$. The condition shows that
the 3-form must be pure spatial volume in the coframe basis, in other words it is determined by a single function, as
\be \label{cdm}
\bM^0 = \frac{1}{2}\rho_{D}\star \bbe^0\,,
\ee 
where we call the function $\rho_{D}$ and the factor $1/2$ is just conventional. 

\subsubsection{The $\bM^a$ describes conserved energy.}

Now it is already clear that if interpreted as some effective material energy-momentum current, the 3-form $\bM_a$ describes a pure energy 
current in the sense that it does not contain an effective pressure component. Taking a pull-back to a spatial hypersurface, it is clear that  $\bM^0$ in
(\ref{cdm}) can be interpreted as an energy of a 3-space volume element, and we can write $\bM^0 = (2v)^{-1}M \bdiff x\wedge\bdiff y\wedge\bdiff z$, 
where $M$ is the mass of a unit coordinate volume $v$. Note that $M/(v\rho_D) = \sqrt{\det{h}}$ can be identified as the determinant of the 3-space metric. 
The time-component of the conservation equation (\ref{cdma}) shows that the energy is 
constant, 
\be
\bDiff\bM_0 = \bdiff \bM_0 - \bomega^I{}_0\wedge\bM_I = \frac{1}{2v}\dot{M}(\bdiff t\wedge \bdiff x\wedge\bdiff y\wedge\bdiff z)= \frac{M'}{2v\sqrt{\det{h}}}\volume = 0\,. 
\ee 
The prime denotes the reparameterisation-invariant time derivative $M'=\partial M/\partial\phi=\dot{M}/\dot{\phi}$ wrt the khronon.
A region of space $\mathcal{V}$ is associated with an energy $\int_\mathcal{V} \bM^0$ which does not change as the region $\mathcal{V}$ expands, contracts or changes its shape. This is
exactly how ideal dust would behave. In particular, material CDM would consist of massive particles which dilute in an expanding universe
such that the energy density is simply inversely proportional to the volume. The fluid approximation would break down when probing
so small scales that the collisions and other possible interactions between the particles would have to be taken into account. This distinguishes
material and the geometrical realisations of CDM. In the present case, the fluid description of CDM is not approximate
but exact. The $M$ is not the mass of a particle contained in a volume $v$ but the mass of the space that spans a volume $v$. 

In summary, we have proven that (given $\m\bO^{ab}=0$ as follows from (\ref{action})), the 3-form $\bM^a$ introduces only one integration constant
into the solutions of the theory, and this constant determines the density of an effective ideal dust which interacts only gravitationally. Therefore it is an
obvious candidate for the missing mass of the universe. 

\subsection{The $\Lambda$}
\label{lambda}

It is well-known that in unimodular gravity the cosmological constant $\Lambda$ appears as a constant of integration. 
Exploited in general relativity since the 1910's, the unimodular device continues to find interesting new applications in current
research, e.g. \cite{Jirousek:2020vhy,Girelli:2021pol,Magueijo:2021rpi,Carballo-Rubio:2022ofy}.  Covariant formulations of unimodular gravity 
\cite{Henneaux:1989zc} reveal that the 3-space is a carrier of information about time \cite{Baierlein:1962zz} cf. Misner volume time,
and below we show that this is compatible with the Lorentz gauge theory.

The action formulation we consider is
\be \label{Laction}
I_\Lambda = \int \Lambda\lp\bdiff\bkappa-\volume\rp\,,
\ee 
where the new fields are the scalar $\Lambda$ and the 3-form $\bkappa$ which is the Lagrange multiplier that sets the constancy of the cosmological constant. We note that
in an alternative formulation, the dark matter candidate 3-form $\bM^0$ as well can be understood as the Lagrange multiplier (which determines the coframe as the derivative of the khronon). The EoMs for the 2 fields in the action (\ref{action}) are
\bs
\ba
\bdiff\Lambda = 0\,, \\
\bdiff\bkappa = \volume\,.
\ea
\es
We shall integrate the previous equation over a 4-volume $\mathcal{W}$ bounded by two Cauchy surfaces $\mathcal{V}_1$ and $\mathcal{V}_2$,
\be \label{4volume}
\int_\mathcal{W} \volume= \int_\mathcal{W} \bdiff\bkappa = \int_{\mathcal{V}_2}\bkappa -  \int_{\mathcal{V}_1}\bkappa\,. 
\ee 
Thus, the $\Lambda$-conjugated time between $\mathcal{V}_2$ and $\mathcal{V}_1$ is the invariant spacetime volume enclosed by these hypersurfaces. 
On the other hand, the invariant lapse of the CDM-conjugated khronon time $\Delta\phi = \phi_2-\phi_1$ is determined by another fundamental field of the theory.
We conclude that there is a relation between the two concepts of time and therefore only one of them can be chosen arbitrarily. An explicit example will be checked below at (\ref{kk}) in \ref{isodyn}. 

Is there a duality between the conjugates of the two concepts of time as well? Can we determine the energy scales of the integration constants $\Lambda$ and $M$ from \1 principles, or at least fix one of them given the other? It is at this point that the answers are ``yet tentative'' as the abstract disclaimed. Perhaps, the local and the global views of time could be related to what Dirac referred to as the atomic units and the Einstein units, in the context of a profound idea of a connection between the cosmological evolution and the constants of Nature known as the large number hypothesis \cite{doi:10.1098/rspa.1974.0095}, which though has not yet found its precise and viable mathematical expression. In the similar way that the khronogenesis gives rise to the constant $c$, the speed of light, the breaking of the de Sitter to the Lorentz can give rise to the gravitational constant \cite{Koivisto:2021ofz}. Finally, the Planck constant could be the
result of the reduction of the conformal into the de Sitter symmetry. The 3-form $\bkappa$ can be related to the kairon scalar field $\kappa^a=\ast\bkappa\lrcorner\bDiff\phi^a$ which is the dual of our symmetry-breaking field $\phi^a$ in a $SO(6,\mathbb{C})$ extension of the $SO(4,\mathbb{C})$ Lorentz gauge theory \cite{Koivisto:2019ejt}. The duality suggests $\Lambda\bkappa \sim \bM_a\phi^a$ which leads to the solution of the $\bkappa \sim \Lambda^{-1}Mt\ast \bdiff t$, in a coordinate system with time $t$. It is plausible that the two fundamental scales 
of cosmology could be predicted in a more complete theory containing the Lorentz gauge theory.   


\section{Cosmology}

We shall now consider the cosmology of the Lorentz gauge theory. In \ref{isokin} we construct the isotropic and homogeneous geometry, and in \ref{isodyn} we derive the dynamical equations and study their exact solution in a simple toy model. The purpose is to demonstrate that the Lorentz gauge theory can provide a completion of the standard inflationary hot big bang cosmology.

\subsection{Isotropic and homogeneous kinematics}
\label{isokin}

The point of departure is the generic isotropic and homogeneous, spatially flat Ansatz for the fundamental fields,
the Lorentz gauge potential 1-form $\bomega^a{}_b$ and the khronon scalar field $\phi^a$ (for a spatially curved case, see \cite{Koivisto:2022uvd}). The generic Ansatz is given by 3 independent functions of time,
\bs
\label{ansatz}
\ba
\phi^a & = & \delta^a_0\phi(t)\,, \\
\bomega^I{}_0 & = & A(t)\bdiff x^I\,, \\
\bomega^I{}_J & = & iB(t)\epsilon^I{}_{JK}\bdiff x^K\,.  
\ea
\es
We may split the connection also as
\bs
\ba
\prescript{\pm}{}\bomega^I{}_0 & = & \frac{1}{2}\lp A \mp B\rp\bdiff x^I\,, \label{comega1} \\
\prescript{\pm}{}\bomega^I{}_J & = & \frac{i}{2}\lp \mp A + B\rp\epsilon^I{}_{JK}\bdiff x^K\,. 
\ea
\es
The coframe field composed from these fields,
\bs
\label{coframe}
\ba
\bbe^0 & = & \dot{\phi}\bdiff t\,, \\
\bbe^I & = & \phi A\bdiff x^I\,,
\ea
\es
shows how the Friedmann-Lema$\hat{\text{i}}$tre geometry emerges from the fields in (\ref{ansatz}). The lapse function $n(t)=\dot{\phi}$ is the time derivative of the khronon. The scale factor $a(t)=\phi A$ is obtained by multiplying the khronon\footnote{For this reason, the choice of the clock $\phi=a$ is not justified from the more fundamental theory, though apparently natural from the perspective of the constructed geometry. One can easily check that the clock  $\phi=a$ is only available in a universe with the curvature-dominated equation of state.} with the component of the gauge potential $A$. The role of the component $B$ we have to deduce from the dynamical equations. 

First we shall look at the gauge-invariant characterisation of the geometry in terms of the field strengths.
The field strength (\ref{curvature}) for the Ansatz (\ref{ansatz}) is
\bs
\label{cosmoR}
\ba
\bR^I{}_0 & = & \dot{A}\bdiff t\wedge \bdiff x^I - i AB\epsilon^I{}_{JK}\bdiff x^J\wedge\bdiff x^K\,, \\
\bR^I{}_J & = & i\dot{B}\epsilon^I{}_{JK}\bdiff t\wedge \bdiff x^K + \lp A^2 + B^2 \rp \bdiff x^I \wedge \bdiff x_J\,. 
\ea
\es
From this we obtain, according to (\ref{Rab}), the 2-forms
\bs
\label{FandT}
\ba
\bF^I & = & -\frac{i}{2}\lp \dot{A}-\dot{B}\rp \bdiff t\wedge\bdiff x^I + \frac{1}{4}\lp A - B \rp^2\epsilon^I{}_{JK}\bdiff x^J\wedge\bdiff x^K\,, \label{F} \\
\bT^I & = & \phi\lp \dot{A}\bdiff t\wedge\bdiff x^I - iAB\epsilon^I{}_{JK}\bdiff x^J\wedge\bdiff x^K\rp\,. \label{T}
\ea
\es
It is a useful cross-check to verify that $\bF^I$ is the Ashtekar curvature according to (\ref{ashtekarcurvature}). The metric spin connection in cosmology is given by
\bs
\label{connections}
\ba
\mathring{\bomega}^I{}_0 & = & \dot{a}/\dot{\phi}\bdiff x^I\,, \\
\mathring{\bomega}^I{}_J & = & 0\,, 
\ea
as one readily deduces from (\ref{coframe}). On the other hand, from (\ref{omega+}) and (\ref{comega1}) we see that now the extrinsic curvature is
\be
\bK^I = (A-B)\bdiff x^I = 2i\bA^I\,.
\ee
\es
With the simple expressions (\ref{connections}) plugged into the formulae (\ref{ashtekarcurvature}) both of them result in (\ref{F}), so the cross-check is passed and we may continue
to the field equations.

\subsection{Isotropic and homogeneous dynamics}
\label{isodyn}

Since we have introduced various different connections along the way, we can consider the different torsions wrt these connections. For the cosmological Ansatz (\ref{ansatz}), the torsion of the self-dual and the antiself-dual, the Levi-Civita and the Ashtekar connections are, respectively,
\bs
\label{torsions}
\ba
\prescript{\pm}{}\bT^I & = & \lb \phi\dot{A} + \frac{1}{2}\dot{\phi}\lp A \pm B\rp \rb\bdiff t\wedge\bdiff x^I + \frac{i}{2}\phi A\lp \pm A -B\rp \epsilon^I{}_{JK}\bdiff x^J\wedge\bdiff x^K\,,  \\
\mathring{\bT}^I & = & 0\,, \label{sdtorsion} \\
\boldsymbol{\mathcal{T}}^I & = & \lp\phi \dot{A} + \dot{\phi}A \rp \bdiff t\wedge\bdiff x^I + \frac{i}{2}\phi A\lp A-B\rp\epsilon^I{}_{JK}\bdiff x^J\wedge\bdiff x^K\,.  
\ea
\es
It is illuminating to revisit the field equations (\ref{efe2}) in their fully covariant form before adapting them to cosmology in the time gauge. Exploiting the geometric identity (\ref{torsioncurvature}), we see that the 2 equations can be rewritten as
\bs
\label{efe10}
\ba
i\+\bDiff\+\bT^a & = & \bt^a + \bM^a\,, \\
i\bDiff\bB^{ab} & = & \bO^{ab}\,. 
\ea
\es
This makes the structure of the theory quite transparent: the self-dual torsion is the excitation whose flux is sourced by energy momentum, and the (proto)area element $\bB^{ab}=\+\bDiff\phi^{[a}\wedge\bDiff\phi^{b]}/2$ is the excitation sourced by angular momentum,
\bs
\label{area}
\ba
\bB^{0I} & = & \frac{1}{4}\phi A\lp \dot{\phi}\bdiff t\wedge\bdiff x^I + \frac{i}{2}\phi A\epsilon^I{}_{JK}\bdiff x^J\wedge\bdiff x^K\rp\,, \\
\bB^{IJ} & = & -\frac{1}{4}\phi A\lp\dot{\phi}\epsilon^{IJ}{}_K\bdiff t\wedge\bdiff x^K + \frac{1}{2}\phi A \bdiff x^I\wedge\bdiff x^J\rp\,. 
\ea
\es
In section \ref{genesis} we saw that in the Minkowski limit the self-dual torsion will vanish, consistently with our interpretation of $\+\bT^I$ as the material energy excitation. Eq. (\ref{sdtorsion}) shows that when there is no expansion, $\dot{a}=0$, either $A=B$ must be the same constant or otherwise $A=0$. As we learned in \ref{genesis}, the angular excitation $\bB^{ab}$ does not necessarily vanish in the absence of material sources. 

To complete the cosmological system, we need to specify the matter sources. We assume an isotropic and homogeneous perfect fluid. Such a source is completely determined by
its energy density $\rho_M$, pressure $p_M$ and angular momentum $\Omega$, and these may only depend on time. To first recover the standard Friedmann equations, we consider 
the case of vanishing angular momentum, $\Omega=0$. Then $\bO^{ab}=0$ and the energy current is determined as
\bs
\label{sources}
\ba
\bt^0 & = & \frac{1}{2}\rho_M \star\bbe^0\,, \label{source1} \\
\bt^I & = & -\frac{1}{2}p_M\star\bbe^I\,,  \label{source12}
\ea
\es 
Plugging these sources into the above equations (\ref{efe10}) using the torsion of the self-dual connection (\ref{sdtorsion}) and the area excitation (\ref{area}), or alternatively plugging the sources into the equations (\ref{summaryeqs}) formulated in terms of the 2-forms which are given by (\ref{FandT}), we arrive at the 3 equations 
\bs
\label{generalfriedmann}
\ba
3\lp A-B\rp^2 A\phi & = & Mv^{-1} + \rho_M\lp A\phi\rp^3\,, \label{eka} \\
2\lp \dot{A} - \dot{B}\rp A\phi + \lp A-B\rp^2\dot{\phi} & = & -p\lp A\phi\rp^2\dot{\phi}\,, \label{toka} \\
\lp \dot{A}\phi + B\dot{\phi}\rp A\phi & = & 0\,. \label{vika}
\ea
\es
The system is trivially solved by the ground state $\phi=0$, as well as the Minkowski solution $A=0$ which implies also $B=0$, and therefore we will assume $A\phi \neq 0$.  Eq.(\ref{vika}) is then solved by $B=-A'\phi$, where the prime denotes derivative wrt the khronon. Plugging into (\ref{eka}) gives the \1 Friedmann equation
\bs
\label{friedmann}
\be
3a(a')^2 = M v^{-1} + a^3\rho_M \quad \Rightarrow \quad 3H^2 = \rho_D + \rho_M\,, 
\ee
and (\ref{toka}) yields the \2 Friedmann equation
\be
 2 aa'' + \lp a' \rp^2 = -a^2 p_M  \quad \Rightarrow \quad 2H' + 3H^2 = -p_M\,. 
\ee
\es
As expected, we recover the standard cosmological dynamics in general relativity. The $H$ in (\ref{friedmann}) is defined as the reparameterisation-invariant expansion rate, 
\be
H = \frac{\dot{a}}{na} = \frac{\dot{a}}{\dot{\phi}a} = (\log{a})'\,,
\ee 
which is why the lapse function $n$ and the rate $\dot{n}/n$ do not explicitly enter into the Friedmann equations. 

It is well-known that cosmological trajectories are generically traced back to a singularity as one extrapolates them
back in time (e.g. \cite{Hawking:1966sx,Borde:2001nh}). The question arises whether the fundamental fields of the Lorentz gauge theory could remain well-behaved in the limit that the composite scale factor $a=\phi A \rightarrow 0$ vanishes, and the expansion rate $H$ of this composite scale factor and its higher time derivatives such as $H'$ hit infinity. If it makes sense to talk about the beginning of the universe, the only natural beginning is the symmetric phase $\phi=0$. Anything else would be something rather than nothing. It seems that the connection coefficients $A$ and thus also $B$ could smoothly evolve across $\phi=0$ without any obvious obstacles. However, the invariant description of the gauge field dynamics is in terms of the field strengths. 
In the case at hand, recovering the standard cosmological solution in general relativity for $a \neq 0$, the 2-forms (\ref{FandT}) read
\bs
\ba
\bF^I & = & -\frac{i}{2}a'' \bdiff \phi\wedge\bdiff x^I + \frac{1}{4}\lp a' \rp^2 \epsilon^I{}_{JK}\bdiff x^J\wedge\bdiff x^K\,, \label{grfrw}\\
\bT^I & = & \lp a' - \frac{a}{\phi}\rp\lp \bdiff \phi\wedge\bdiff x^I + i a  \epsilon^I{}_{JK}\bdiff x^J\wedge\bdiff x^K\rp\,.
\ea
\es
Let us assume that the energy density of the primordial universe is dominated by a source with some constant equation of state $w=p/\rho$. The solution to the Friedmann equations
(\ref{friedmann}) then determines us the scalings of the coefficients of the above 2-forms,
\bs
\label{s}
\ba
a & \sim & \phi^{\frac{2}{3(1+w)}}\,, \label{s1} \\
a' \sim a/\phi & \sim & \phi^{\frac{-(1+3w)}{3(1+w)}}\,,  \label{s2} \\
\lp a' \rp^2 & \sim & \phi^{\frac{-2(1+3w)}{3(1+w)}}\,,  \label{s3} \\
a''  & \sim & \phi^{\frac{-2(2+3w)}{3(1+w)}}\,.  \label{s4}
\ea
\es
Whilst (\ref{s1},\ref{s2},\ref{s3}) stay finite as $\phi \rightarrow 0$ for accelerating equation of state, $-1<w<-1/3$ (effectively, violation of the strong energy condition), (\ref{s4})
requires the stronger condition $-1<w<-2/3$. The scaling (\ref{s4}) describes the electric component of the self-dual field strength (\ref{grfrw}). The Lorentz gauge theory thus
{\it predicts inflation from khronogenesis} since a decelerating universe cannot appear from nothing. 

The action is not only stationary $\delta I=0$ but realises the density-free boundary condition $\bL\lvert_{\phi=0}=0$ since ground state does not contribute to the action\footnote{Therefore the action is stationary wrt arbitrary variations of the fields \cite{Koivisto:2022oyt}. Thus the conclusions do not hinge on Dirichlet or any other boundary conditions at the initial time $\phi=0$. Nothing changes if the integration limit is extended towards $\phi \rightarrow -\infty$, since there is no action there, $\bL=0$.}. 
The point $H=\infty$ where the metric curvature $\mathring{\bR}{}^a{}_b$ diverges in general relativity and other conventional formulations of gravity is understood as
the totally symmetric phase in the ground state $\phi=0$ of the Lorentz gauge theory which is consistent with {\it any} gauge field strength ${\bR}^a{}_b$. 
The density and the pressure are of course divergent $\rho_M \sim p_M \sim \phi^{-2}$, but they are not directly observable nor fundamental fields of the theory, but quantities derived by dividing physical charges by spatial volumes. The matter Lagrangian is expected to scale as $\bL_M \sim \phi^{-2w/(1+w)}$, and thus continuously disappears as we wind the clock back to $\phi=0$. 
The constant-$w$ toy model thus predicts not only something rather than nothing, but something material in an initially inflating background.   
In particular, as seen already from the solution we derived in \ref{genesis}, time cannot begin in an empty flat space. Note that a bouncing scale factor is not suggested, since that would require $w<-1$ (or the violation of the weak energy condition). 


To close this section, we check how cosmic dual time evolves according to considerations in \ref{lambda}.   
It is easy to solve (\ref{4volume}) using (\ref{s}). If we consider a unit chunk of space until the khronon time $\phi$, the LHS gives the spacetime volume
\bs
\label{kk}
\be
\int \star 1  = \frac{1+w}{3+w}\phi^{\frac{3+w}{1+w}}v\,.
\ee  
On the other hand, the LHS gives, assuming a homogeneous and isotropic 3-form $\bkappa$,
\be
\int\bkappa = \phi^{\frac{2}{1+w}}\kappa v \quad \Rightarrow \quad \kappa = \frac{1+w}{3+w}\phi\,.
\ee 
\es
Thus, the kairon time $\kappa$ is simply proportional to the khronon time $\phi$ in the case of the constant-$w$ solution. The de Sitter solution $w=-1$ is a limiting case for which $\kappa$ is the constant $\kappa=1/3H$.  


\section{Cosmology with spin}

Material spin currents, being the Noether currents corresponding to Lorentz symmetry, are of a paramount interest in Lorentz gauge theory.
It is not obvious that the generalised Friedmann equations (\ref{generalfriedmann}) are consistent with nontrivial dynamics in the presence of a spin current $\Omega$. 
Its presence will presumably modify the energy conservation of the matter sources, besides modifying the gravitational dynamics.
Below in \ref{conservation} we derive the general consistency conditions for spinning fluids in the Lorentz gauge theory by studying its Noether identities, and
then in \ref{spinfluid} we apply the conditions to the case of a perfect spinning fluid in isotropic and homogeneous cosmology. A class of exact solutions is presented.

\subsection{Spin and energy conservation}
\label{conservation}

We had parameterised the variation of the matter action by introducing the currents in (\ref{mattervariations}). More precisely, when the matter fields $\psi$ obey their 
Euler-Lagrange equations, we have
\be \label{deltaLM}
\delta \bL_M = \bdiff\lp -\bt_a\delta\phi^a + \delta\psi\wedge\frac{\partial L_M}{\partial \bDiff\psi}\rp + \delta\phi^a\bDiff\bt_a + \delta\bomega^{ab}\wedge\lp \phi_{[a}\bt_{b]} - \bO_{ab}\rp\,. 
\ee  
The density $\bL_M$ is taken to be invariant $\delta_\lambda \bL_M=0$ under the infinitesimal Lorentz transformation with parameters $\lambda^a{}_b$,
\bs
\label{lorentz}
\ba
\delta_\lambda \phi^a & = & \lambda^a{}_b\phi^b\,, \\
\delta_\lambda \bomega^a{}_b & = & -\bDiff\lambda^a{}_b\,.
\ea
\es
We may consider parameters $\lambda^a{}_b$ s.t. that they vanish at the boundary, in which case we can neglect the symplectic piece in (\ref{deltaLM}) as well as another boundary term
which arises from a partial integration after inserting (\ref{lorentz}) in (\ref{deltaLM}). We then see that the Lorentz transformation is an invariance of the density $\bL_M$ if 
\be \label{lorentznoether}
\bDiff\bO^{ab} = \bDiff\phi^{[a}\wedge\bt^{b]}\,.
\ee
This identity holds even off-shell since the field equations were not used in the derivation. We have reproduced the result that the divergence of the spin tensor is the antisymmetric energy tensor.

The matter action should also be invariant under infinitesimal diffeomorphisms parameterised by a vector $\xi$. The symmetry is exact only up to a boundary term $\delta_\xi \bL_M = \xi\lrcorner\bL_M$, 
but the boundary term is not relevant for the Noether identity. Diffeomorphisms in Lorentz gauge theory
generate spacetime geometry from the fundamental fields. The latter are the khronon and the gauge potential and the former, as we have seen, is constructed in terms of the (co)frame
and the curvature. So,
\bs
\label{diffeo}
\ba
\delta_\xi \phi^a & = & \xi\lrcorner\bDiff\phi^a\,, \\ 
\delta_\xi \bomega^a{}_b & = & \xi\lrcorner\bR^a{}_b\,.
\ea
\es
We adapt the generic variation (\ref{deltaLM}) to this case (\ref{diffeo}) and again neglect a boundary term. Choosing the vector $\xi=\bbie_a$ to be one of the four ``legs'' of the Vierbein,
the resulting Noether identity assumes the form
\bs
\label{diffnoether}
\be
\bDiff\bt_a = -\bbie_a\lrcorner\bR^{bc}\wedge\lp\phi_{[b}\bt_{c]} - \bO_{bc}\rp\,.  \label{diffnoether1}
\ee
Matter fields without spin (or nonminimal couplings) are described by energy currents without divergence wrt the metric connection $\mathring{\bDiff}\bt_a=0$, regardless of the gravity theory \cite{Koivisto:2005yk}. It is useful to recover this from our result (\ref{diffnoether1}). 
For this purpose, let us decompose the gauge potential $\bomega^a{}_b = \mathring{\bomega}^a{}_b + \bC^a{}_b$ in terms of the Levi-Civita connection $ \mathring{\bomega}^a{}_b\wedge\bbe^b = -\bdiff\bbe^a$ and the contorsion $\bC^a{}_b\wedge\bbe^b = \bT^a$. Using this decomposition we can rewrite (\ref{diffnoether1}) as
\be
\mathring{\bDiff} \bt_a = -(\bbie_a\lrcorner\bC^b{}_c)\bt_b\wedge\bDiff\phi^c + \bbie_a\lrcorner\bR^{bc}\wedge\bO_{bc}\,. \label{diffnoether2}
\ee
By taking into account the previous Noether identity (\ref{lorentznoether}) we can rewrite this in yet another form
\be
\mathring{\bDiff} \bt_a = (\bbie_a\lrcorner\bC^{bc})\bDiff\bO_{bc} + \bbie_a\lrcorner\bR^{bc}\wedge\bO_{bc}\,. \label{diffnoether3}
\ee
\es
which makes it manifest that the usual covariant conservation law for material energy currents only has to be generalised for spinning matters in the Lorentz gauge theory. 

\subsection{Isotropic and homogeneous spin fluid}
\label{spinfluid}

We are now ready to derive the spin fluid conservation equations in cosmology. In particular, we consider the Ansatz (\ref{sources}) for the energy current, and now 
take into account also the spin current\footnote{The cosmological implications of Weyssenhoff spinning perfect fluids have been investigated in the context of Poincar{\'e} gauge theories of gravity, e.g. \cite{PhysRevLett.56.2873,Obukhov:1987yu,Brechet:2008zz}. The Weyssenhoff 
fluid obeys the so called Frenkel condition (see though e.g. \cite{Boehmer:2006gd,Iosifidis:2020gth} for alternatives), with the interpretation that the spin reduces to a pure rotation in the rest frame of the fluid. 
The equivalent form of the cosmological Ansatz (\ref{asource}), not subject to the Frenkel condition $\bO^{ab}\bbie_a\lrcorner\bDiff\phi^0=0$, has also been considered in Poincar{\'e} gauge theory cosmology, e.g. \cite{Kranas:2018jdc,Akhshabi:2023xan}. From the derivations below one can confirm that (\ref{asource}) correctly describes spin (and not boost) in the Lorentz gauge theory, since it excites the magnetic (and not the electric) sector of geometry.}
\bs
\label{asource}
\ba
\bO^{0I} & = & \frac{1}{2}\Omega\star\bbe^I\,,   \label{source3} \\
\bO^{IJ} & = & -\frac{i}{2}\Omega\epsilon^{IJ}{}_K\star\bbe^K\,.  \label{source4}
\ea
\es
We note that the cosmological symmetry would allow to consider the 2 independent functions $\Omega$ and $\tilde{\Omega}$ for the 2 sets of components of the spin current (\ref{asource}). However,
the field equation would enforce $\tilde{\Omega}=\Omega$ since they only are compatible with a self-dual spin current $\bO^{ab}=\+\bO^{ab}$. This we assumed already in the Ansatz (\ref{asource}) because it is the consequence of the minimal coupling of matter fields $\psi$ to the self-dual connection in the action (\ref{action}).

The generalised Friedmann equations (\ref{generalfriedmann}) now read, in terms of the gauge-invariant time variable, 
\bs
\label{generalfriedmann2}
\ba
3\lp A-B\rp^2 a & = & M v^{-1} + \rho_M a^3\,, \label{ekas} \\
2\lp {A}' - {B}'\rp a + \lp A-B\rp^2 & = & -p_M a^2\,, \label{tokas} \\
(\log{A})' a + B & = & \Omega a\,. \label{vikas}
\ea
\es
Before using these field equations, we will consider the spin current (\ref{asource}) in light of the off-shell Noether identities.

First, it is easy to see that there is no antisymmetric energy source for the divergence of the spin, since according (\ref{sources}) we have that $\bt^a \sim \star\bbe^a$ and therefore
\be
\bbe^{[a}\wedge\bt^{b]} \sim  \bbe^{[a}\wedge\star\bbe^{b]} = -\eta^{[ab]}\star 1 = 0\,. 
\ee
The LHS of the (\ref{lorentznoether}) also consistently vanishes as it is easy to check using (\ref{asource}). Thus, the Noether identity resulting from the Lorentz 
invariance is trivially satisfied in the isotropic and homogeneous setting.

We then compute the diffeomorphism invariance Noether identity (\ref{diffnoether}). There identity has 1+3 Lorentz components, but the 3 space components vanish trivially and we can focus on the 1 time component. We begin with the LHS of (\ref{diffnoether3}), recalling the metric connection from (\ref{connections}), and arrive at
\bs
\label{results}
\be
\mathring{\bDiff}\bt_0 = \frac{1}{2}\lb \rho_M' + 3H\lp \rho_M + p_M\rp \rb \star 1\,. 
\ee 
We have already deduced that the \1 term in the RHS of (\ref{diffnoether3}) vanishes
\be
(\bbie_0\lrcorner\bC^{bc})\bDiff\bO_{bc} = -(\bbie_0\lrcorner\bC^b{}_c)\bt_b\wedge\bDiff\phi^c = 0\,,
\ee
and do not actually need the contorsion coefficients $\bC^I{}_0 = (A-aH)\bdiff x^I$ and $\bC^I{}_J = i B\epsilon^I{}_{JK}\bdiff x^K$ in this computation. The \2 term in the RHS of (\ref{diffnoether3}) we obtain
by plugging in (\ref{cosmoR},\ref{asource}) into
\be
-\bbie_0\lrcorner\bR^{bc}\wedge\bO_{bc} = -\frac{3}{a}\lp A' - B'\rp \Omega\star 1\,. 
\ee
\es
Combining the results (\ref{results}) we obtain
\bs
\label{continuity}
\be 
\rho_M' + 3H\lp \rho_M + p_M\rp = -\frac{6}{a}\lp A' - B' \rp \Omega\,, \label{ekakont}
\ee
from the time component of the diffeomorphism Noether identity (\ref{diffnoether}).

Let us now go on shell. Taking the time derivative of the \1 Friedmann eq. (\ref{ekas}) and then using the \2 Friedmann eq. (\ref{tokas}) we arrive at
\be
\rho_M' + \lp 2H + \frac{A-B}{a}\rp \rho_M + \frac{3\lp A- B\rp}{a} p_M = 0\,, 
\ee
where the spin function $\Omega$ is not explicitly involved. However, we can yet use the \3 Friedmann eq. (\ref{vikas}) to solve $(A-B)/a=H-\Omega$, and the above equation
becomes
\be
\rho_M' + 3H\lp \rho_M + p_M\rp = \lp \rho_M + 3p_M\rp\Omega\,.  
\ee
\es
Using the \1 and \2 Friedmann equations to rewrite the RHS of (\ref{ekakont}) in terms of the matter density and pressure, we confirm that all the forms (\ref{continuity})
are on-shell equivalent. Thus, this is the consistent generalisation of the matter continuity equation in the presence of cosmological spin fluid. The corresponding result has been
derived recently in the context of Poincar{\'e} gauge cosmology \cite{Kranas:2018jdc}.

The generalised Friedmann equations (\ref{generalfriedmann}) are reduced to  
\bs
\ba
3\lp H - \Omega\rp^2 & = &  \rho_D + \rho_M\,, \\
2\lp H' - \Omega'\rp + \lp 3H -\Omega\rp \lp H-\Omega\rp & = & -p_M\,,
\ea
\es
or equivalently, in terms of an effective energy density $\rho_{\text{eff}}$ and effective pressure $p_{\text{eff}}$,
\bs
\ba
3H^2 & = & \rho_{\text{eff}}  + \rho_D\,, \quad \text{where} \quad \rho_{\text{eff}} = \rho_M + 6H\Omega - 3\Omega^2\,, \\
2H' + 3H^2  & = & -p_{\text{eff}}\,, \quad \text{where} \quad p_{\text{eff}} = p_M - 2\Omega' - 4H\Omega + \Omega^2\,, 
\ea
\es
To solve these equations, the properties of the fluid source have to be specified. Now this requires the 2
equation of state parameters, to determine the spin $\Omega$ as well as the pressure $p_M$. The former has the dimension
$M$ and the latter has the dimension $M^4$, and since the only dimensional quantities we have at hand are $H$ and $\rho_M$,
the apparently natural form of the equations of state would be $\Omega = \alpha H$ and $p_M = w_M \rho_M$ with some
dimensionless parameters $\alpha$ and $w_M$. The simplest assumption is that both these 2 parameters are constant.
Assuming the spin fluid effective energy dominates $a^3v\rho_{\text{eff}} > M$, we can then immediately solve the equations, e.g. by integrating (\ref{continuity}),
and find that the cosmological expansion corresponds to the effective equation of state 
\be \label{weff}
w_{\text{eff}} = \frac{\rho_{\text{eff}}}{p_{\text{eff}}} = w_M\lp 1-\alpha\rp - \frac{\alpha}{3}\,.
\ee
If we consider the very earliest moments of the universe, the fields are expected to be in the radiation-like $w_M=1/3$ phase.
Then any $\alpha>1$ leads to an accelerating universe, $\alpha=2$ corresponding to the de Sitter -like phase with $w_{\text{eff}}=-1$.

This raises the possibility of a completely new kind inflation \cite{PhysRevLett.56.2873}, not driven by the pressure exerted by a hypothetical 
slowly-rolling scalar field but by the spins of ordinary matter fields which, hypothetically, become significant at extremely 
high energies. This could explain not only the inflationary beginning of the universe from khronogenesis but also the smooth recovery
of the standard hot big bang cosmological evolution. (In the conventional scalar field models, one has to device an exit from inflation 
and an accompanying so called reheating process which then transforms the scalar field into ordinary matter fields.)

The $\Omega$ introduces novel possibilities also in the context of dark energy cosmologies. The discrepancy between the
Hubble evolution as deduced from the source density and from the actual one, as characterised by (\ref{weff}), suggests that
the effect of $\Omega$ on the expansion rate could address the $H_0$ tension\footnote{A proof of this concept can already found in the literature \cite{Akhshabi:2023xan}, see also \cite{Poplawski:2019tub,Izaurieta:2020xpk}.}, the statistically significant 
discrepancy of the expansion rate normalisations as inferred from observations at larger and at smaller redshifts within the standard
$\Lambda$CDM model  \cite{DiValentino:2021izs}. For example, a spinning dark energy could in principle account for the dark matter 
phenomenology $w_M=0$ but nevertheless predict $w_{\text{eff}} \approx -0.7$ in the present universe. Again, the condition for acceleration is $\alpha>1$.

 \section{Discussion}

To assess the viability of the scenarios we have discussed in these proceedings, it is necessary to study the effects of spinning fluids on cosmological 
perturbations. At the level of perturbations, further parameters or assumptions will be required to determine the evolution of the source fluid, now for the spin 
as well as for the pressure. In section \ref{spinfluid} we exploited a simple parameterisation at the background level to arrive at exact solutions, and this 
approach can be extended to uncover the phenomenological impact of spin currents in the cosmic microwave background radiation and in the large-scale structure formation. 

However, from the perspective of the Lorentz gauge theory,  the most interesting approach is to develop the Lagrangian formulation of fields with nonzero spin. Novel phenomenology may result without invoking any new exotic ingredients, since indeed all the elementary fields of the standard model of particle physics have spin (with one exception if the Higgs is supposed to be elementary). In conventional gauge theories of gravity, spinors are well-known to induce a 4-fermion interaction via axial torsion \cite{Magueijo:2012ug}, but the implications of the fermionic spin in the Lorentz gauge theory remain to be explored. A consistent coupling of spinor matters $\psi$ to the self-dual connection \cite{Ashtekar:1989ju} would result in
$\bO^{ab} \sim \bar{\psi}(\gamma^c \gamma^{[a}\gamma^{b]}\psi_- +  \gamma^{[a}\gamma^{b]}\gamma^c\psi_+)\star(\bDiff\phi_c)$, where $\gamma^a$ are elements of the Clifford algebra $\gamma^{(a}\gamma^{b)}=-\eta^{ab}$. 
Also, since the standard model gauge fields have spin 1, they can be associated with spin currents. This was found to be possible in Gallagher's pregeometric \1 order Yang-Mills theory, in the presence of an effective vacuum excitation \cite{Gallagher:2022kvv}. Finally, we recall that even the immaterial 3-form $\bM^a $ may have a role in supporting spin currents when the assumption we made in section \ref{aligned} is relaxed.

There is an intriguing relation between the latter two effects. The CDM candidate 3-form $\bM_a$ is the gravitational analogy of vacuum polarisation in Yang-Mills theories. Relaxing the assumption of section \ref{aligned} results simultaneously in the analogy of vacuum magnetisation in the gravitational sector, and the possible spin currents of the gauge fields in the Yang-Mills sector. In such a phase of the theory, the 3-form $\bM_a$ no longer describes ideal dust. Khronogenesis in this phase could therefore be followed by the usual $\sim$60 e-folds of inflation without diluting the energy in the 3-form $\bM_a$ to completely negligible (which would be the result for an ideal dust energy). Some such mechanism is called for if aiming to unify both the suggested rationale for the initial conditions of the universe and the suggested ingredients for a $\Lambda$CDM theory into a more complete paradigm of cosmology based on the Lorentz gauge theory. 


\vskip25pt

\section*{Acknowledgments}

\begin{itemize}
\item The results reported in these proceedings are based on collaborations with Priidik Gallagher, Luca Marzola, Ludovic Varrin and Tom Z\l{}o\'snik.
\item The author would like to thank the organisers and the participants of the workshop Metric Affine Frameworks for Gravity at Tartu 2022.
\item Sections 2,3,4,5 were edited from notes to the lectures ``Physics in Action'' in the series of courses Geometrical Foundations of Gravity.    
\item Thanks to the referee of IJGMMP for the 10 useful suggestions to improve this manuscript. 
\item This work was supported by the Estonian Research Council grant PRG356 ``Gauge Gravity - unification, phenomenology and extensions''.
\end{itemize}

\bibliographystyle{siam}
\bibliography{LCDMrefs}


\end{document}